\documentclass{cernrep} 
\usepackage{texnames}
\usepackage[T1]{fontenc}
\usepackage[bookmarks, colorlinks=true, linktoc=page, pdftex, linkcolor=red, citecolor=red, urlcolor=red]{hyperref}
\usepackage{grffile}
\usepackage{graphicx}
\graphicspath{{./figs/}}
\usepackage{subcaption}
\newcommand{\ab}[1]
{\allowbreak{}}

\begin{document}
\title{Status of MERLIN}
 
\author{H. Rafique$^{1~2~3}$, R. B. Appleby$^{2~3}$, R. J. Barlow$^1$, J. G. Molson$^5$, S. Tygier$^{2~3}$, A. Valloni$^{2~4}$}

\institute{$^1$University of Huddersfield, Huddersfield, UK, $^2$University of Manchester, Manchester, UK, \\$^3$Cockcroft Institute, Daresbury, UK, $^4$CERN, Geneva, Switzerland,\\$^5$LAL, Univ. Paris-Sud, CNRS/IN2P3, Universit\'{e} Paris-Saclay, Orsay, France}

\maketitle

\begin{abstract}
MERLIN is an accelerator physics library written in C++ which can be used for a range of accelerator tracking simulations, including collimation in hadron colliders. Recently MERLIN has been upgraded to provide a robust tool for HL-LHC collimation, including the treatment of composite materials, and a hollow electron lens process. We describe the features of MERLIN used in collimation simulations.
\end{abstract}

 
\section{Introduction}

MERLIN is an accelerator physics library written in C++, created by Nick Walker at DESY in 2000, to study the International Linear Collider~\cite{ILC} (ILC) beam delivery system ground-motion~\cite{merlinilc}. Later, the main linac and damping rings were added,~\cite{merlinilc2} necessitating wakefield, collimation, and synchrotron radiation processes~\cite{merlinwakefield}. As the ILC is an electron linac, the TRANSPORT maps were deemed acceptable for particle tracking. Later, Andy Wolski added synchrotron functionality, including a module to calculate the Courant-Snyder parameters, closed orbit and dispersion, and symplectic integrators for many turn simulations.

The current loss map tool for the LHC is SixTrack~\cite{sixtrack}, a particle tracking code that was updated to include the K2~\cite{K2} scattering routines for collimation~\cite{demolaizesixtrack} and has also been combined with the FLUKA~\cite{fluka}.

It was decided to update MERLIN to include the requirements for a complementary simulation of the LHC and future collimation system. MERLIN is written in C++ making it modular, and easy to modify. It offers thick lens tracking, an on-line aperture check, and a number of physics processes. The scattering physics has recently been updated to include more advanced proton-nucleon elastic and single diffractive scattering~\cite{molson}. These updates, ensure that MERLIN offers a fast, accurate, and future-proofed tool for ultra-relativistic proton tracking, collimation, and a robust hollow electron lens process~\cite{hrthesis}.

MERLIN has been used for simulating loss maps for the existing and future upgrades to the LHC collimation systems~\cite{ipac16_WEPMW036, ipac16_WEPMW037}. The source code is now available on GitHub~\cite{merlingit}, and is described in detail in~\cite{molson} and~\cite{hrthesis}. In this article we describe the features of MERLIN that are used in collimation studies. Dedicated articles on MERLIN's composite materials and hollow electron lens are included in these proceedings~\cite{merlin_cm, merlin_hel}.

\section{MERLIN}

For a typical loss map simulation, MERLIN must be given a set of input files that describe the accelerator lattice, the machine aperture and the collimator settings. Then a beam of particles is created, based on the accelerator optics and distribution requested, and tracked around the lattice. When a particle hits a collimator, the scattering through the material is simulated. Finally, if a particle loses sufficient energy in a scatter, or hits a non collimator component, its loss is recorded and a loss map is generated.

\begin{figure}[!htbp]
	\begin{center}
		\includegraphics[width=0.5\linewidth]{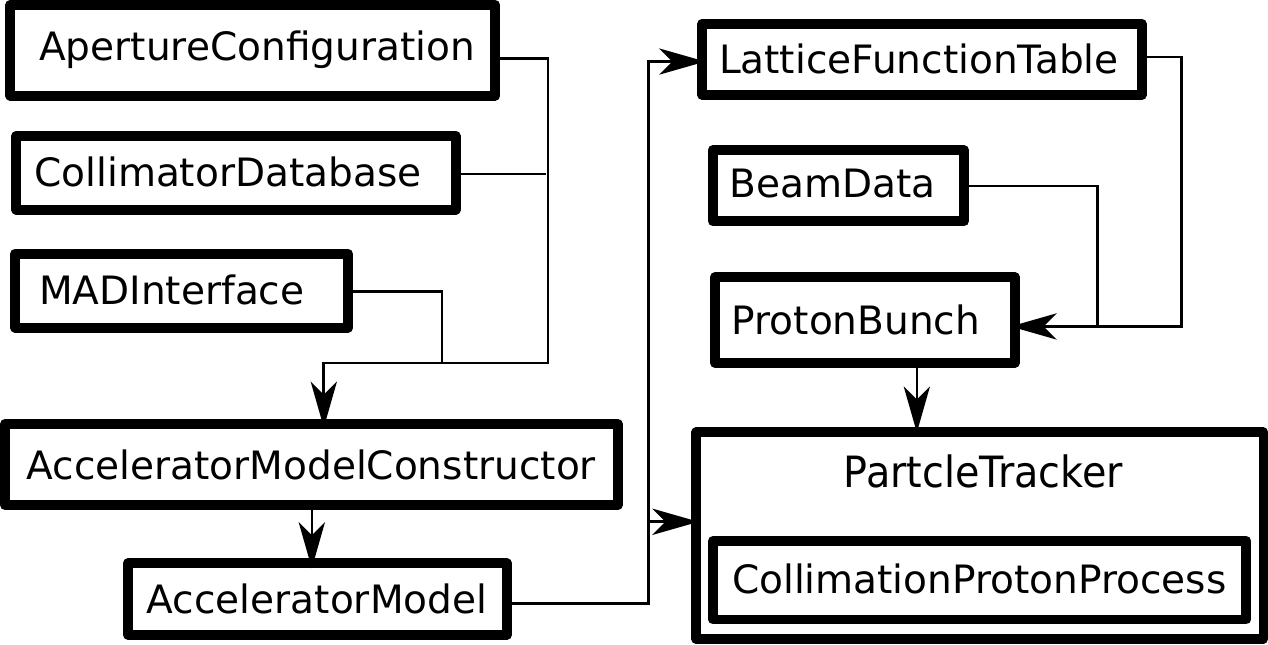}
		\caption{The data flow of a typical collimation simulation using MERLIN}
		\label{fig:codeflow}
	\end{center}
\end{figure}

The user may define their accelerator in the form of an \texttt{Accelerator\ab{}Model}, using the \texttt{MAD\ab{}Interface} class to read a standard MADX~\cite{madx} TFS table. Further input files may be read by the \texttt{Aperture\ab{}Configuration} class to define the apertures of the accelerator, and the \texttt{Collimator\ab{}Database} to set up collimators. The user may then calculate the lattice functions of the accelerator, and in turn use these to define a beam, which leads to the construction of a \texttt{Particle\ab{}Bunch} that is matched to the accelerator at the desired position. A particle tracker may be constructed, selecting from either \texttt{TRANSPORT} or \texttt{SYMPLECTIC} integrator sets, and physics processes may be attached to it. Finally, the user may run the tracking simulation, and create outputs using one of the existing output functions, or define their own. This standard flow of data between MERLIN's main classes for a collimation simulation is shown in \Fref{fig:codeflow}.

\subsection{Lattice setup}

MERLIN stores an accelerator as an \texttt{Accelerator\ab{}Model} object. This requires three input files for the purpose of LHC and HL-LHC collimation, the first for an accurate model of the individual components of the accelerator, the second to define the apertures of the machine, and the third to define collimator parameters. In order to translate these inputs into a MERLIN \texttt{Accelerator\ab{}Model}, a number of input interfaces and configuration utilities are required.

\begin{figure}[!htbp]
	\begin{center}
		\includegraphics[width=1\linewidth]{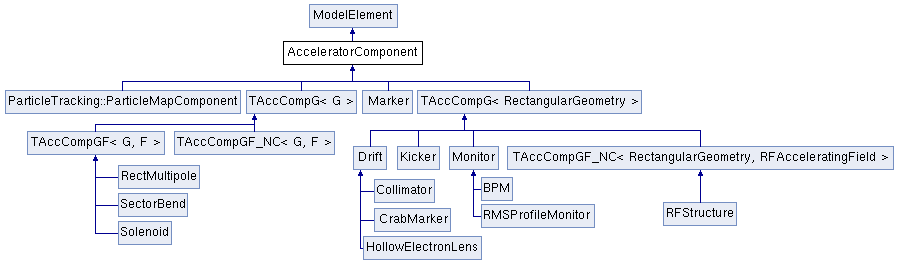}
		\caption{Inheritance diagram of the accelerator components currently available in MERLIN}
		\label{fig:doxyAccComp}
	\end{center}
\end{figure}

The \texttt{Accelerator\ab{}Model} is an ordered vector of \texttt{Accelerator\ab{}Component}s, which is used by the \texttt{Particle\ab{}Tracker} to set the integrators that describe the paths taken by individual particles as they travel through the elements. Each \texttt{Accelerator\ab{}Component} contains an EM field, a geometry, an aperture, and a wake potential object. Special cases exist, for example a \texttt{Collimator} also contains a jaw material which is required for scattering. Figure~\ref{fig:doxyAccComp} shows the inheritance of the standard \texttt{Accelerator\ab{}Component}s (including some internal data types prefixed with \texttt{TAcc}), both \texttt{Collimator} and \texttt{Hollow\ab{}Electron\ab{}Lens} elements are tracked as, and therefore derived from, the \texttt{Drift} component (a vacuum pipe with no field).

A standard MADX thick lens TFS table is passed through the \texttt{MADInterface} class to extract the \texttt{Accelerator\ab{}Component}s. \texttt{MADInterface} reads the column headers so that as long as the required parameters are present, they do not need to be in a fixed order. It iterates through each element and creates the appropriate \texttt{Accelerator\ab{}Component}, appending it to the \texttt{Accelerator\ab{}Model\ab{}Constructor}, a standalone class used in \texttt{MADInterface}. 

MERLIN contains all the standard components required for modelling typical synchrotrons, detailed in Table~\ref{tab:merlincomponents}. Any element may be treated as a drift using the \texttt{MADInterface::\ab{}TreatTypeAsDrift()} function, though this is unwise for certain magnets. MERLIN currently handles a small number of elements as drifts as standard because there is no integrator to perform the expected function, or the expected function cannot be performed using an integrator (for example the hollow electron lens).

\begin{table}[!htbp]
\begin{center}
\caption{Most common accelerator components and their MADX TFS and MERLIN names. For cases such as MULTIPOLE the appropriate element is selected based on the parameters in the lattice. For RBEND, MERLIN uses a SectorBend with appropriate pole face rotations.}
\label{tab:merlincomponents}
\small
\begin{tabular}{|l|c|c|}
\hline
\textbf{Component}	&\textbf{MADX name} &\textbf{MERLIN name} \\
\hline
		Vacuum pipe 				& DRIFT				&	Drift	\\
		RF cavity						& LCAV				&	SWRFStructure	\\
		RF cavity						& RFCAVITY		&	SWRFStructure	\\
		Collimator					& RCOLLIMATOR	&	Collimator	\\
		Collimator					& ECOLLIMATOR	&	Collimator	\\
		Collimator					& COLLIMATOR	&	Collimator	\\
		Rectangular dipole	& RBEND				&	SectorBend	\\
		Sector dipole				& SBEND				&	SectorBend \\
		Multipole						& MULTIPOLE		& Sextupole, Octupole\\
		Vertical corrector	& YCOR				& YCor	\\
		Vertical kicker			& VKICKER			& YCor	\\
		Horizontal corrector& XCOR				& XCor	\\
		Horizontal kicker		& HKICKER			& XCor	\\
		Quadrupole					& QUADRUPOLE	& Quadrupole \\
		Skew quadrupole			& SKEWQUAD		& SkewQuadrupole \\
		Solenoid						& SOLENOID		& Solenoid		\\
		Hollow electron lens& -						& HollowElectronLens \\
		Sextupole						& SEXTUPOLE		& Sextupole \\
		Octupole						& OCTUPOLE		& Octupole \\
		Skew sextupole			& SKEWSEXT		& SkewSextupole \\
		Monitor							& MONITOR			& BPM, RMSProfileMonitor \\
		Marker							& MARKER			& Marker \\
\hline
\end{tabular}
\end{center}
\end{table}

\subsection{Apertures}

MERLIN features on-line aperture checking. Particles' positions are checked against the local aperture at each element as they are tracked through the lattice. The aperture can either be used to stop particles or to trigger the scattering routine in order to simulate the passage through matter.

\begin{figure}[!htbp]
	\begin{center}
		\includegraphics[width=0.7\linewidth]{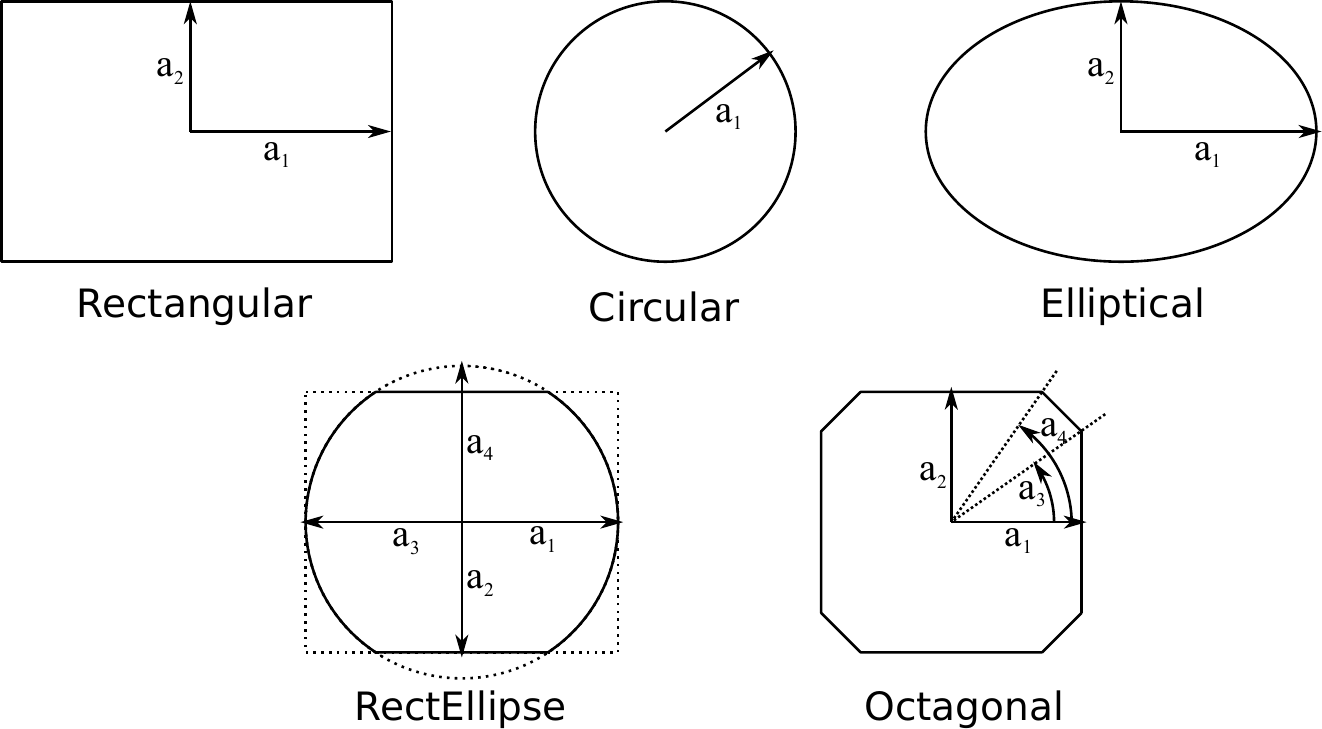}
		\caption{Aperture shapes and geometric parameters}
		\label{fig:apertures}
	\end{center}
\end{figure}

MERLIN contains a number of aperture shapes, as shown in Fig~\ref{fig:apertures}. The basic apertures are defined by up to four geometric parameters, e.g. radius for \texttt{CircularAperture} and half-width and half-height for \texttt{RectangularAperture}. All apertures contain a \texttt{PointInside()} function, which takes the spatial coordinates of a particle $(x,y,z)$, and checks if those coordinates are inside the aperture.

MERLIN provides the \texttt{InterpolatedAperture} set of classes, which allow the geometric parameters to vary along the length of an element. This is useful for describing the beam pipe around the interaction points in the LHC.

For collimator scattering, MERLIN has the \texttt{CollimatorAperture} family of classes. During collimation, if the particle hits the aperture in an element that is a \texttt{Collimator} and has a \texttt{CollimatorAperture}, then instead of being immediately lost, it is passed to a scattering routine. This collimator scattering is described in section \ref{sec:col}.

The apertures for a lattice can be read in from a TFS file using the \texttt{ApertureConfiguration} class. This reads an input table containing the shape and geometric parameters of the apertures, and attaches the appropriate aperture classes to the lattice elements.

\subsection{Particle Bunch}

MERLIN uses three coordinate pairs, $(x,x'),(y,y'),(ct,\delta)$, to define a particle as a six dimensional vector $\textbf{p}$, as shown in Equation \ref{eqn:MERLINp}. $x$ and $y$ are the transverse horizontal and vertical coordinates, $x'$ and $y'$ are the corresponding angles, $\delta$ is the longitudinal energy offset, and $ct$ is the displacement from the reference position in $s$.

\begin{equation} \label{eqn:MERLINp}
	\textbf{p} = \left(
		\begin{matrix}
		x \\
		x' \\
 		y \\
		y' \\
		ct \\
		\delta \\
 		\end{matrix}
	 \right).
\end{equation}

MERLIN stores particles as \texttt{PSVector}s, a class that contains the particle coordinate vector as components, as well as a number of other variables, all of which are detailed in Table.~\ref{tab:psvector}.

\begin{table}[!htbp]
\begin{center}
\caption{Components of the \texttt{PSVector} class}
\label{tab:psvector}
\begin{tabular}{|c|c|c|}
\hline
\textbf{Accessor}	& \textbf{Component}	&\textbf{Index} 	\\
\hline
x			&	$x$									&	0	\\
xp			&	$x'$								&	1	\\
y			&	$y$									&	2	\\
yp			&	$y'$								&	3	\\
ct			&	$ct$								&	4	\\
dp			&	$\delta$							&	5	\\
type		&	Type of last particle scatter		&	6	\\
s			&	Location in lattice					&	7	\\
id			&	Individual particle ID				&	8	\\
sd			&	Single diffractive flag				&	9	\\
\hline
\end{tabular}
\end{center}
\end{table}

The \texttt{Particle\ab{}Bunch\ab{}Constructor} class is used to create an initial bunch matched to the machine lattice functions at any chosen injection position. In order to do this a \texttt{Beam\ab{}Data} object must be created and fed to the bunch constructor.

\texttt{Beam\ab{}Data} provides a data structure for definition of the 6D beam phasespace. Using the \texttt{Lattice\ab{}Function\ab{}Table} the user may define parameters at the injection position, which are fed to the \texttt{Particle\ab{}Bunch\ab{}Constructor}. The components of \texttt{Beam\ab{}Data} are shown in Table~\ref{tab:beamdata}.

\begin{table}[!htbp]
\begin{center}
\caption{Components of the \texttt{Beam\ab{}Data} class. $^1$ Must be specified. $^2$ units are dependent on the type of distribution selected in the \texttt{Particle\ab{}Bunch\ab{}Constructor}. }
\label{tab:beamdata}
\begin{tabular}{|c|c|}
\hline
\textbf{Component(s)}	& \textbf{Parameter(s)}	\\
\hline
x0, xp0, y0, yp0, ct0, dp0					&	Beam centroid$^1$ \\
beta\_x, beta\_y, alpha\_x, alpha\_y		&	Lattice functions$^1$ \\
emit\_x, emit\_y							&	Emittances$^1$ \\
sig\_dp										&	Relative energy spread \\
sig\_z										&	Bunch length \\
p0											&	Reference momentum$^1$ \\
c\_xy, c\_xyp, c\_xpy, c\_xpyp				&	x-y coupling \\
Dx, Dxp, Dy, Dyp							&	Dispersion$^1$ \\
charge										&	Particle charge$^1$ \\
min\_sig\_x, max\_sig\_x					&	Minimum and maximum beam size in x (in $\sigma$)$^2$ \\
min\_sig\_y, max\_sig\_y					&	Minimum and maximum beam size in y (in $\sigma$)$^2$ \\
min\_sig\_z, max\_sig\_z					&	Minimum and maximum bunch length$^2$ \\
min\_sig\_dp, max\_sig\_dp					&	Minimum and maximum energy deviation$^2$ \\
		
\hline
\end{tabular}
\end{center}
\end{table}

MERLIN provides a number of bunch distributions, all of which are stored in the \texttt{ParticleBunchConstructor}. The majority are described in~\cite{molson}. When using the constructor the bunch is matched to the lattice functions at the position of creation. The user may specify the construction of the bunch via an input file.

\begin{figure}[!htbp]
	\begin{center}
		\includegraphics[width=1\linewidth]{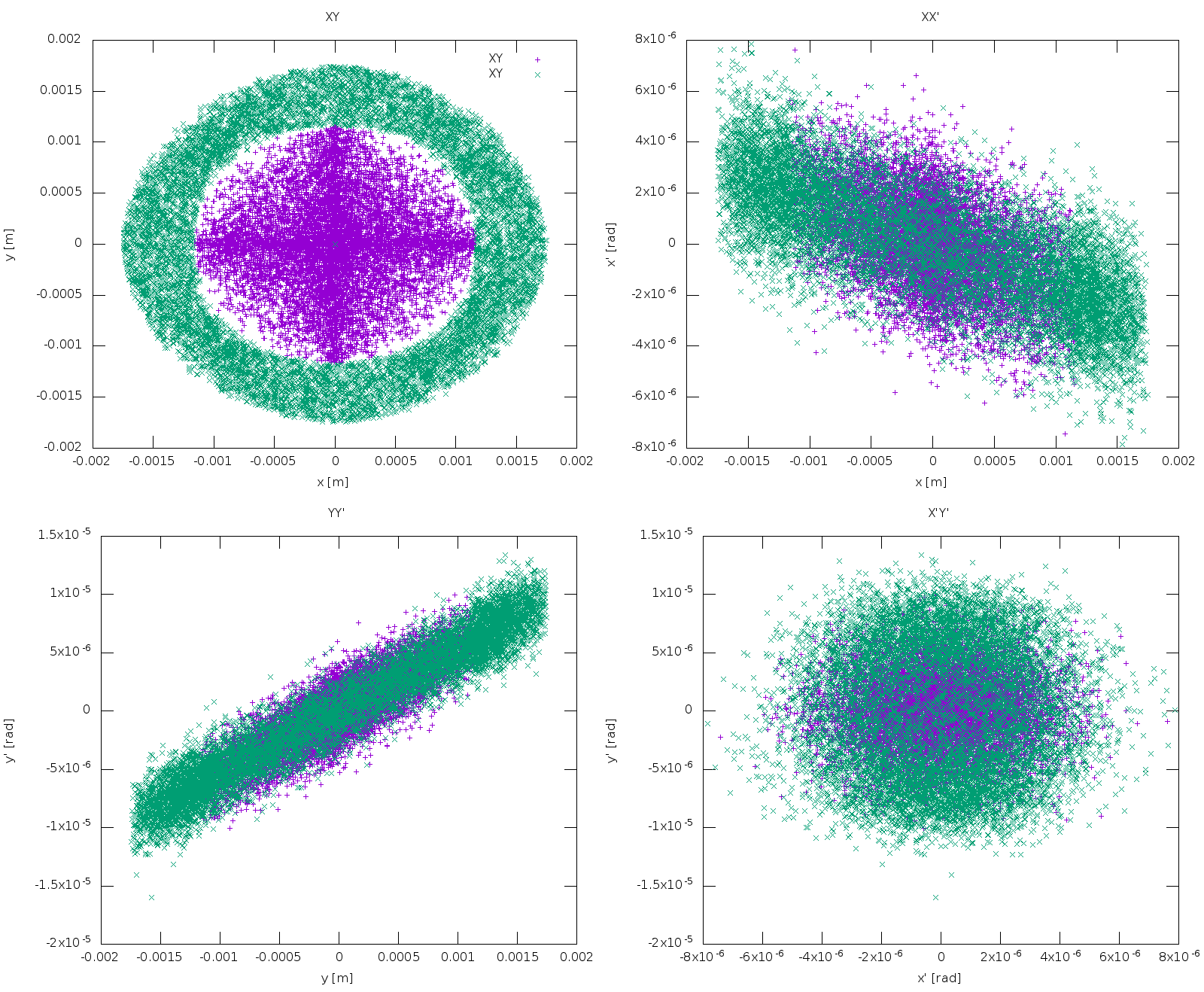}
		\caption{HEL halo distribution in $xy$, $xx'$, $yy'$, and $x'y'$ phasespace. Purple points are a `core' bunch populated between 0-4 $\sigma_x$ and $\sigma_y$, green points are a `halo' bunch populated between 4-6 $\sigma_x$ and $\sigma_y$. This bunch is created at an `injection' position of HEL in the nominal LHC.}
		\label{fig:HELHaloDistn}
	\end{center}
\end{figure}

For the study of the HEL in the LHC and HL-LHC, a HEL halo distribution was created. This is a simple halo bunch that is populated between $\sigma_{x_{min}}$ and $\sigma_{x_{max}}$ in $x$, and $\sigma_{y_{min}}$ and $\sigma_{y_{max}}$ in $y$, thus a matched halo distribution in $xy$ phasespace is generated, as shown in Fig.~\ref{fig:HELHaloDistn}. The minima and maxima are specified using the \texttt{BeamData} class as discussed previously. All other coordinates are matched to regular beam parameters and the optics of the machine at the point of injection in the simulation. 

\subsection{Integrators}

Tracking of particles through each lattice element is performed by integrators. An integrator takes the vector of coordinates of each particle at the beginning of an element and transforms it to the coordinates at the end. A specific integrator is used for each element type and an integrator set will contain an integrator for each common element.

MERLIN contains multiple inbuilt integrator sets including \texttt{TRANSPORT}, which uses second order transport maps and \texttt{SYMPLECTIC}, which defines symplectic integrators for each element. These can be selected using the \texttt{Set\ab{}Integrator\ab{}Set()} method of the tracker. It is also possible to add additional integrators or override individual integrators within a set.

\subsection{Synchrotron Motion}

\begin{figure}[!htbp]
	\begin{center}
		\includegraphics[width=0.8\linewidth]{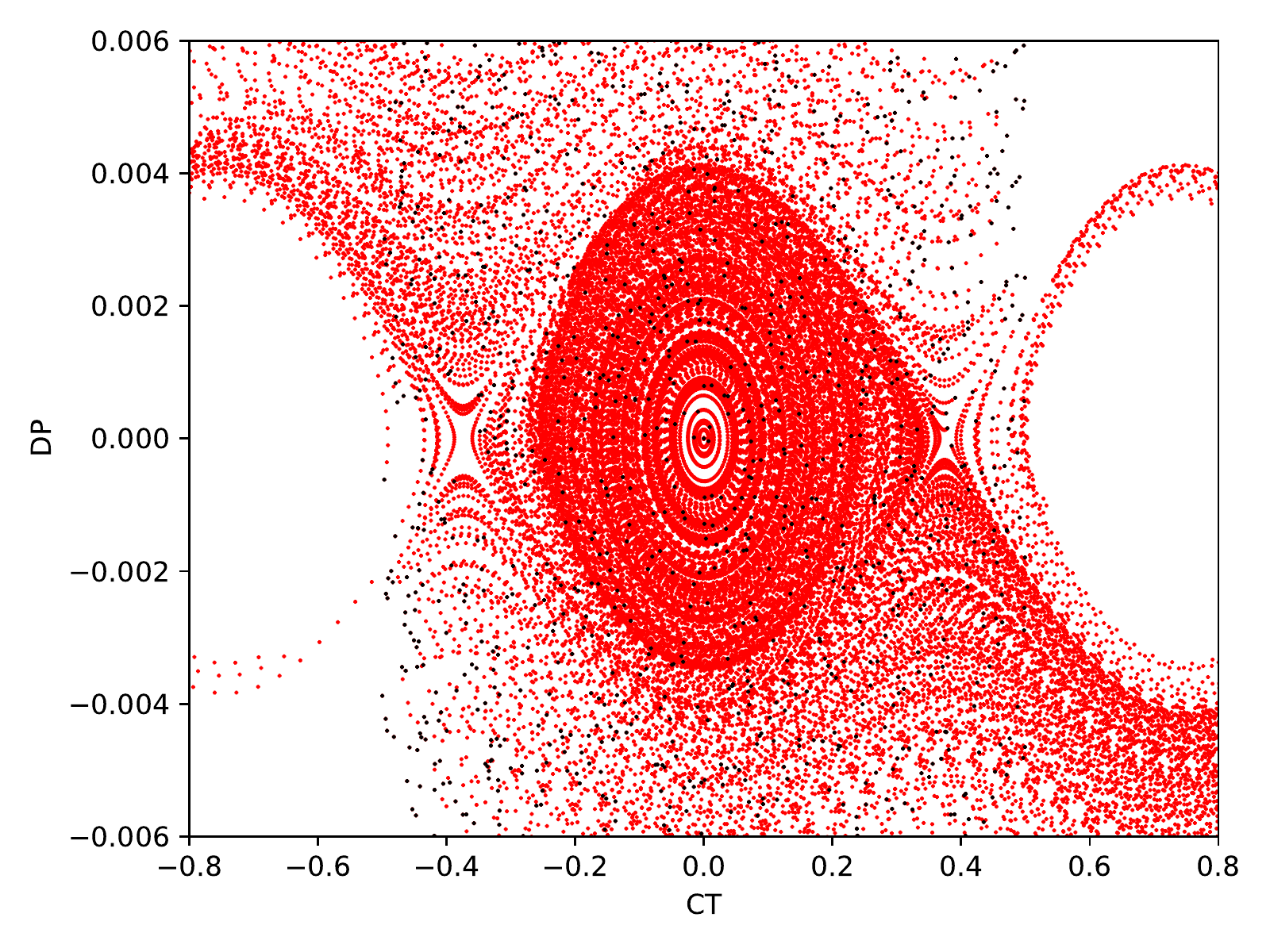}
		\caption{Poincar\'{e} section in $ct, \delta$ phasespace of a large initial distribution (black) over 100 turns in the LHC (red), showing RF bucket and synchrotron motion.}
		\label{fig:RFnocoll}
	\end{center}
\end{figure}

Radiation damping and beam acceleration provide another mechanism for particle loss.
The off-momentum collimation insertion in IR3 of the LHC is designed for this purpose. The RF bucket, and synchrotron motion in MERLIN is demonstrated in Fig.~\ref{fig:RFnocoll}, which shows a Poincar\'{e} section, the phasespace over multiple turns, from an LHC simulation with collimation disabled. 

\begin{figure}[!htbp]
	\begin{center}
		\includegraphics[width=0.8\linewidth]{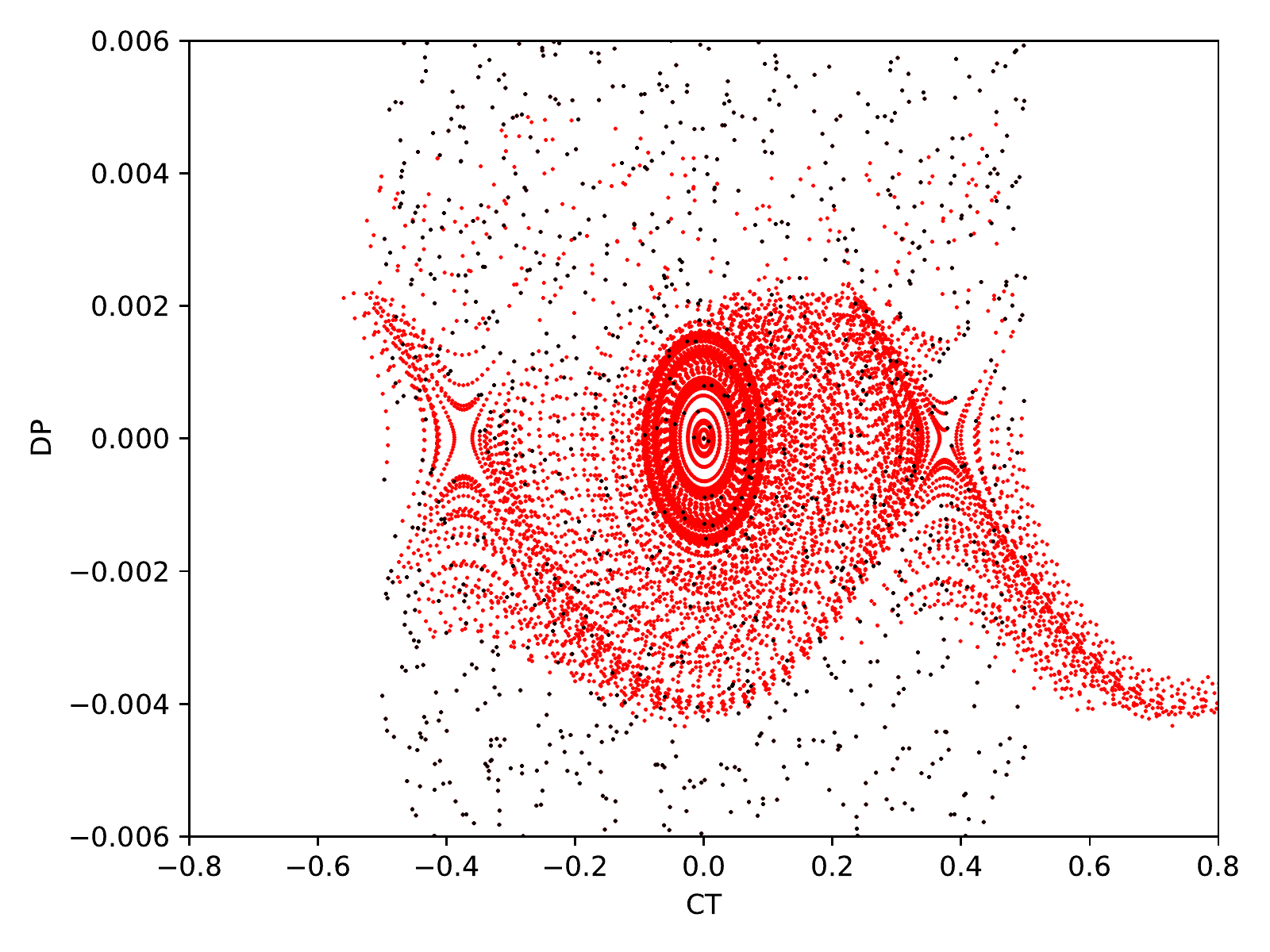}
		\caption{Poincar\'{e} section in $ct, \delta$ phasespace of a large initial distribution (black) over 100 turns in the LHC (red) with collimation enabled and the IR7 transverse TCPs in place.}
		\label{fig:RFcoll7}
	\end{center}
\end{figure}

\begin{figure}[!htbp]
	\begin{center}
		\includegraphics[width=0.8\linewidth]{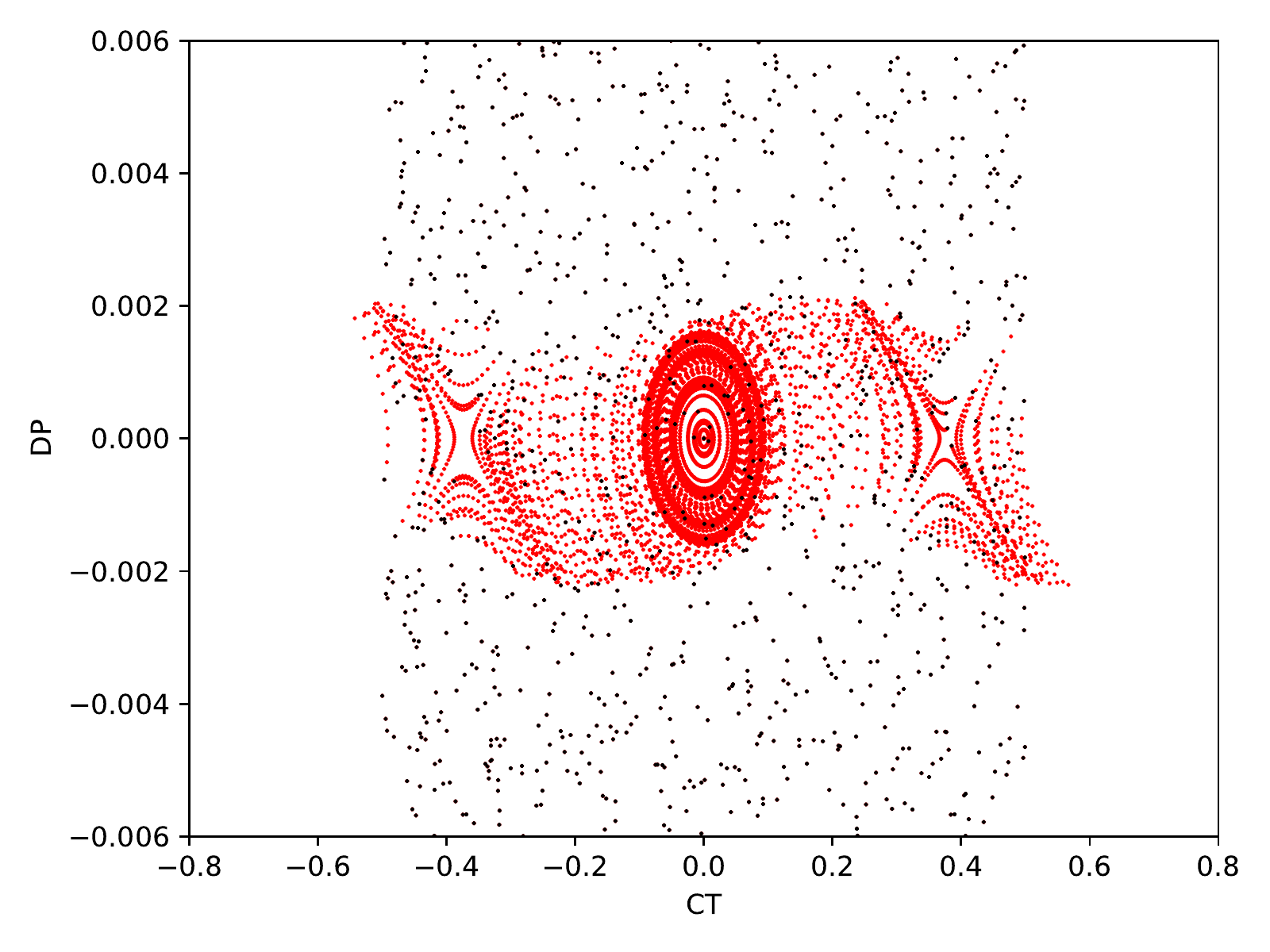}
		\caption{Poincar\'{e} section in $ct, \delta$ phasespace of a large initial distribution (black) over 100 turns in the LHC (red), with collimation enabled and both the IR7 transverse and IR3 longitudinal TCPs in place.}
		\label{fig:RFcoll73}
	\end{center}
\end{figure}

Figures~\ref{fig:RFcoll7} and \ref{fig:RFcoll73} show the same synchrotron motion, but now with collimation enabled, first with only the IR7 transverse primary collimators, and then with both IR7 and the IR3 longitudinal primary collimators. It can be seen that the IR3 collimation region can be used to tightly control energy spread.

\subsection{Lattice Functions}

The Courant-Snyder parameters, along with the closed orbit and dispersion, give a description of the beam envelope and linear optics around the accelerator lattice. They are useful for confirming that the simulation code has a correct model of the accelerator lattice and accurate modelling of particle dynamics. They are also needed for setting up collimator jaw positions and initial beam parameters.

For collimator jaw openings we use units of $\sigma$, which is proportional to the RMS beam emittance and beta function on the given plane at the requested position in the lattice. For example in the $x$ plane $\sigma_x = \sqrt{\beta_x \epsilon_x}$ (when dispersion is zero). This means that when setting collimator apertures we require the lattice functions. Changes in optics -- for example the beta squeeze at the experiments -- cause a change in the position of the collimator jaws.

MERLIN calculates the lattice functions by tracking particles, rather than by the transfer matrix methods found in optics codes. The \texttt{Lattice\ab{}Function\ab{}Table} class takes the \texttt{Accelerator\ab{}Model} and beam energy to first find the closed orbit and then calculate the lattice functions.

The closed orbit is found iteratively: a set of particles with small offsets in each phasespace coordinate is tracked through the lattice and a transfer matrix is calculated from the final coordinates. From this an approximate closed orbit is found, and a new iteration is performed around it. This is repeated until the closed orbit converges.

The lattice functions are then found by again tracking a set of particles with offsets through the lattice, recording their positions after every element. From these coordinates the lattice functions at each element can be calculated.

\begin{figure}[!htbp]
   \centering
   \includegraphics*[width=0.8\columnwidth]{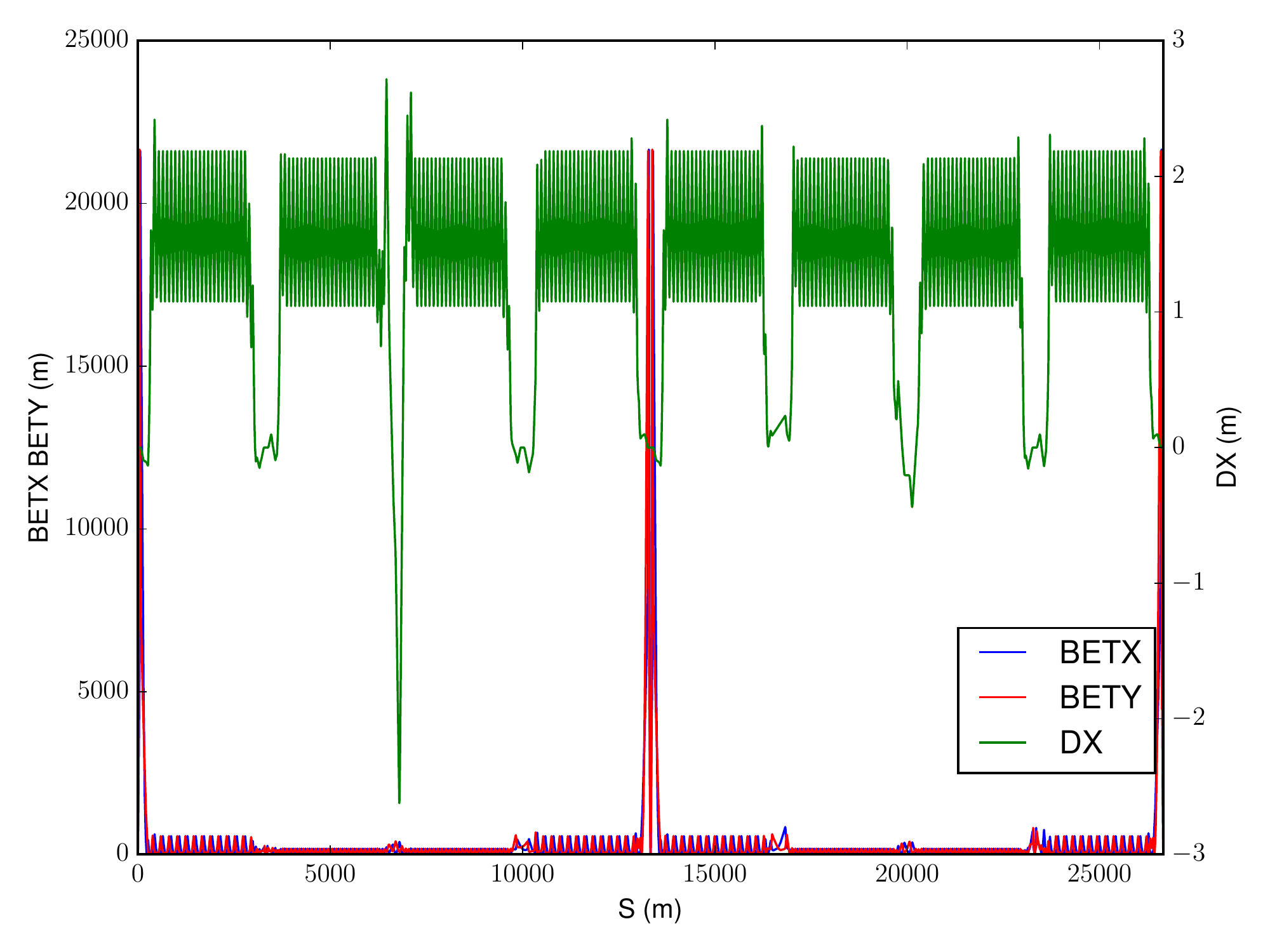}
   \caption{$\beta$-functions and dispersion for HL-LHC ring, 15 cm round optics}
   \label{fig:beta_disp_Ring_BETX_BETY}
\end{figure}

\begin{figure}[!htbp]
    \centering
    \begin{subfigure}[b]{0.49\columnwidth}
        \includegraphics[width=\columnwidth]{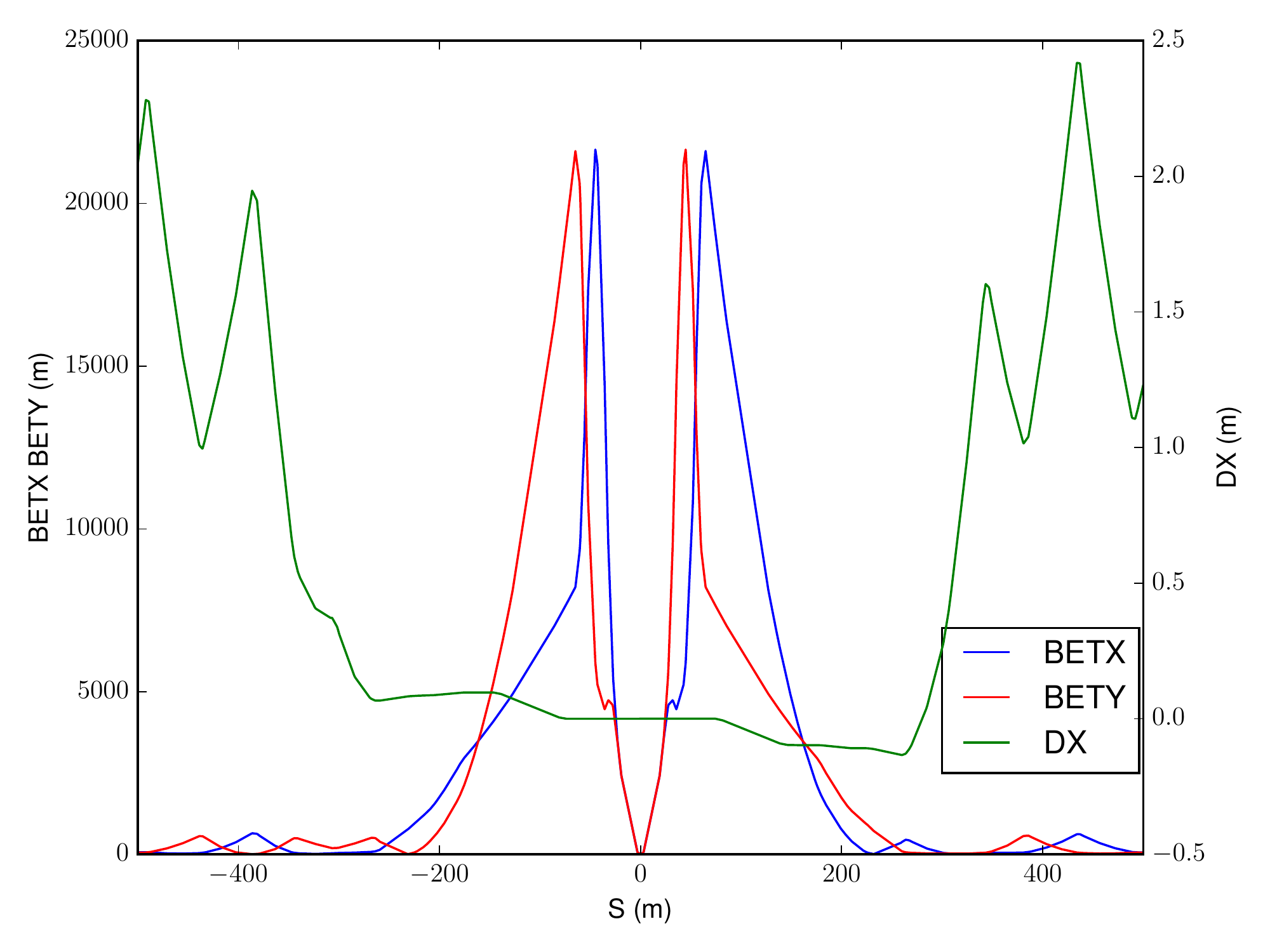}
        \caption{ATLAS IR1}
        \label{fig:beta_disp_ATLAS_IR1_BETX_BETY}
    \end{subfigure}
        \begin{subfigure}[b]{0.49\columnwidth}
        \includegraphics[width=\columnwidth]{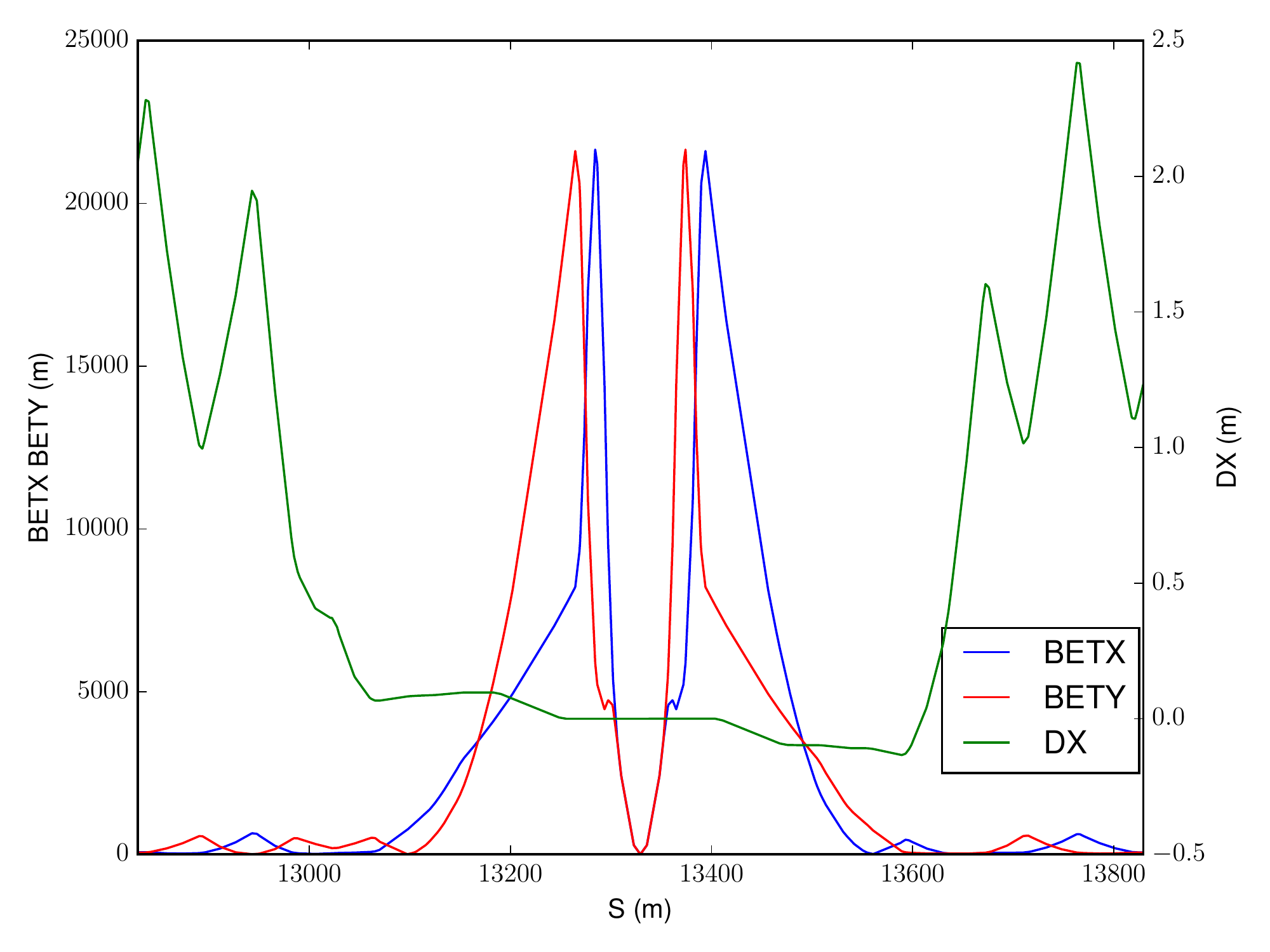}
        \caption{CMS IR5}
        \label{fig:beta_disp_CMS_IR5_BETX_BETY}
    \end{subfigure}
    
        \begin{subfigure}[b]{0.49\columnwidth}
        \includegraphics[width=\columnwidth]{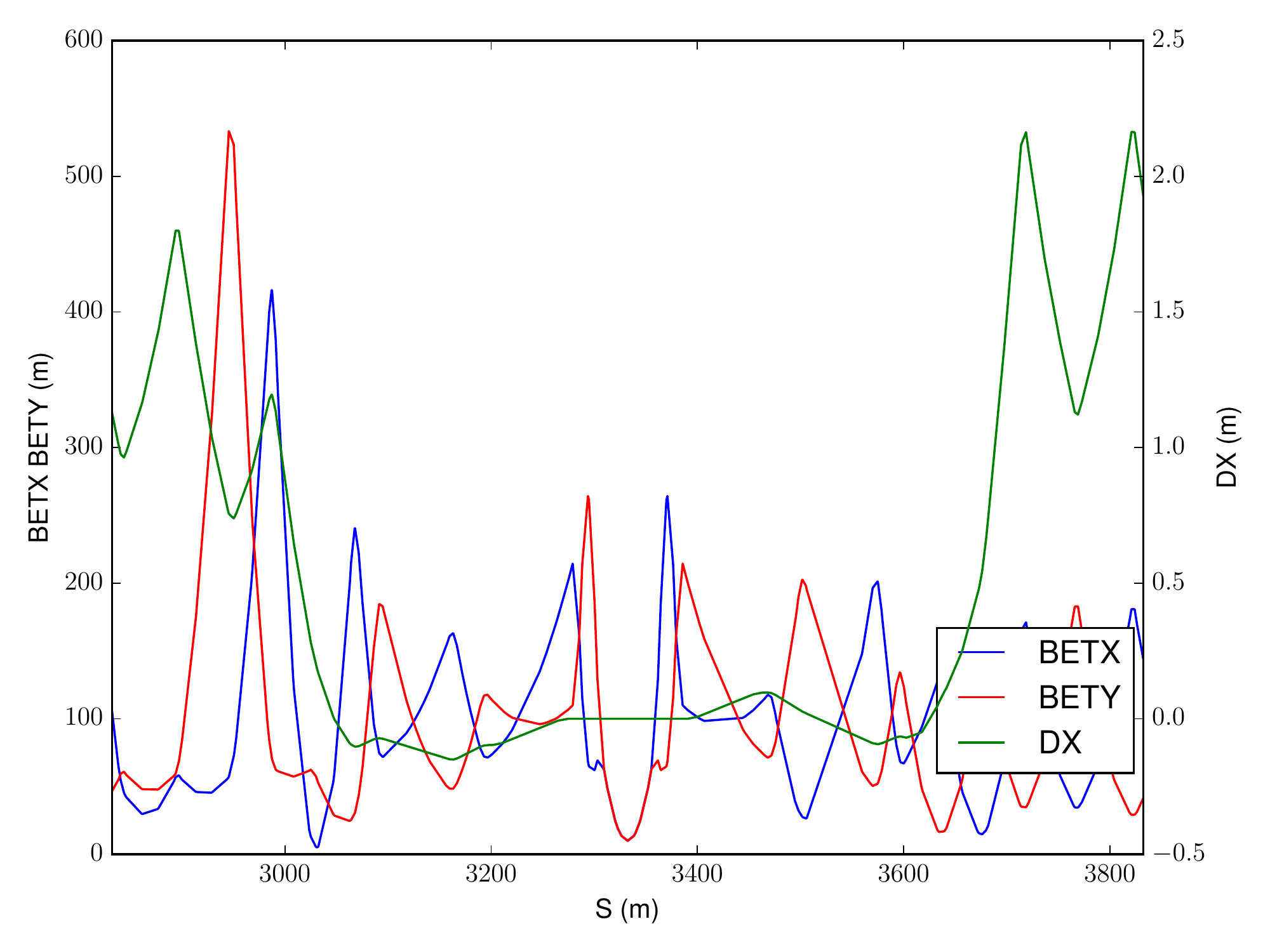}
        \caption{ALICE IR2}
        \label{fig:beta_disp_ALICE_IR2_BETX_BETY}
    \end{subfigure}
        \begin{subfigure}[b]{0.49\columnwidth}
        \includegraphics[width=\columnwidth]{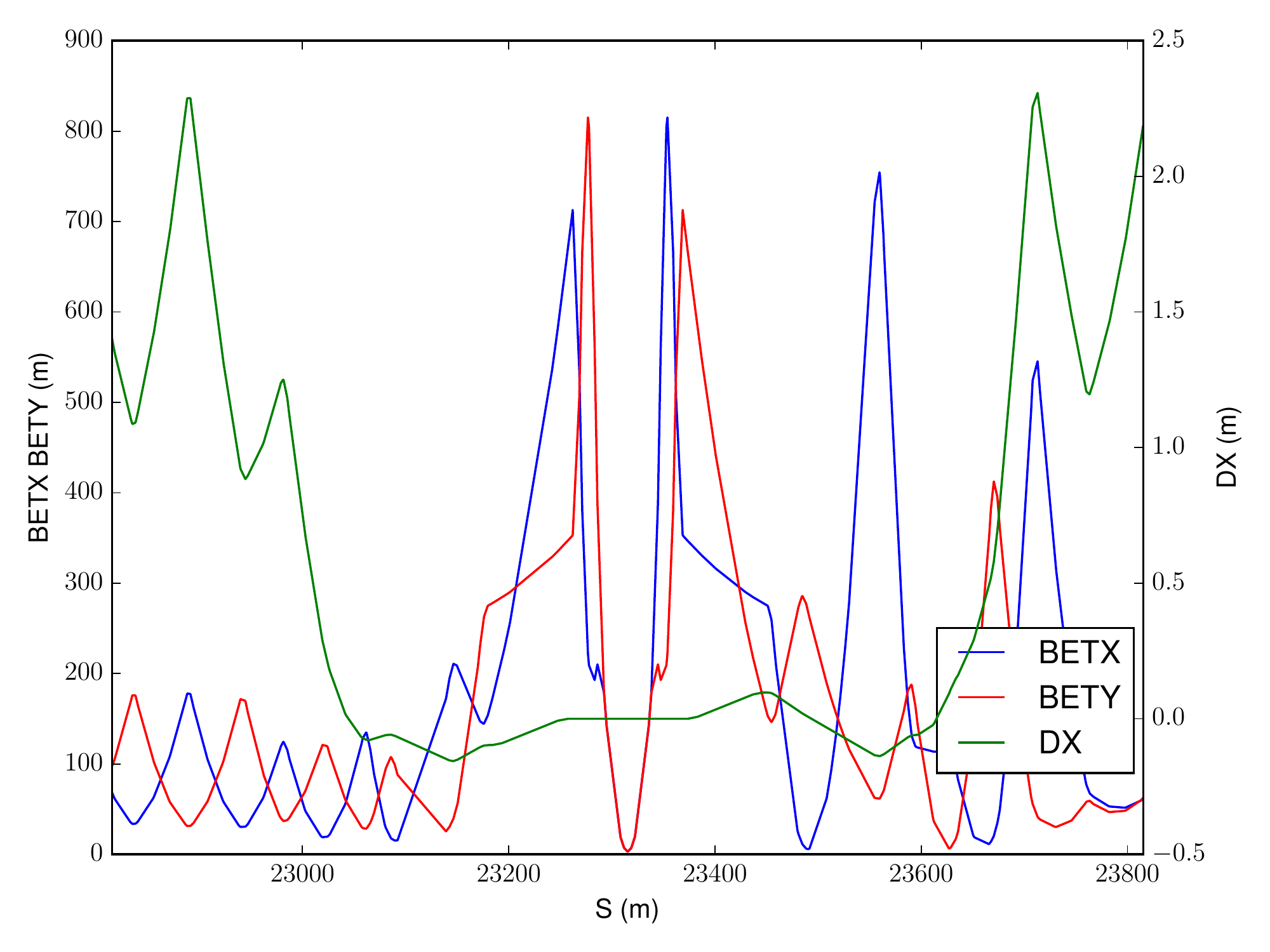}
        \caption{LHCb IR8}
        \label{fig:beta_disp_LHCb_IR8_BETX_BETY}
    \end{subfigure}
    
    \caption{$\beta$-functions and dispersion for HL-LHC experiments, 15 cm round optics}
    \label{fig:beta_disp_IR_BETX_BETY}
\end{figure}

Figure \ref{fig:beta_disp_Ring_BETX_BETY} shows the $\beta$ function and horizontal dispersion around the full ring for the round 15~cm squeezed HL-LHC optics. The increased $\beta$ in the arc adjacent to the high luminosity experiments due to the ATS optics is clearly visible. Figure \ref{fig:beta_disp_IR_BETX_BETY} shows zoomed views of the $\beta$ function and horizontal dispersion at each of the 4 experiments.

\begin{figure}[!htbp]
    \centering
    \begin{subfigure}[b]{0.49\columnwidth}
        \includegraphics[width=\columnwidth]{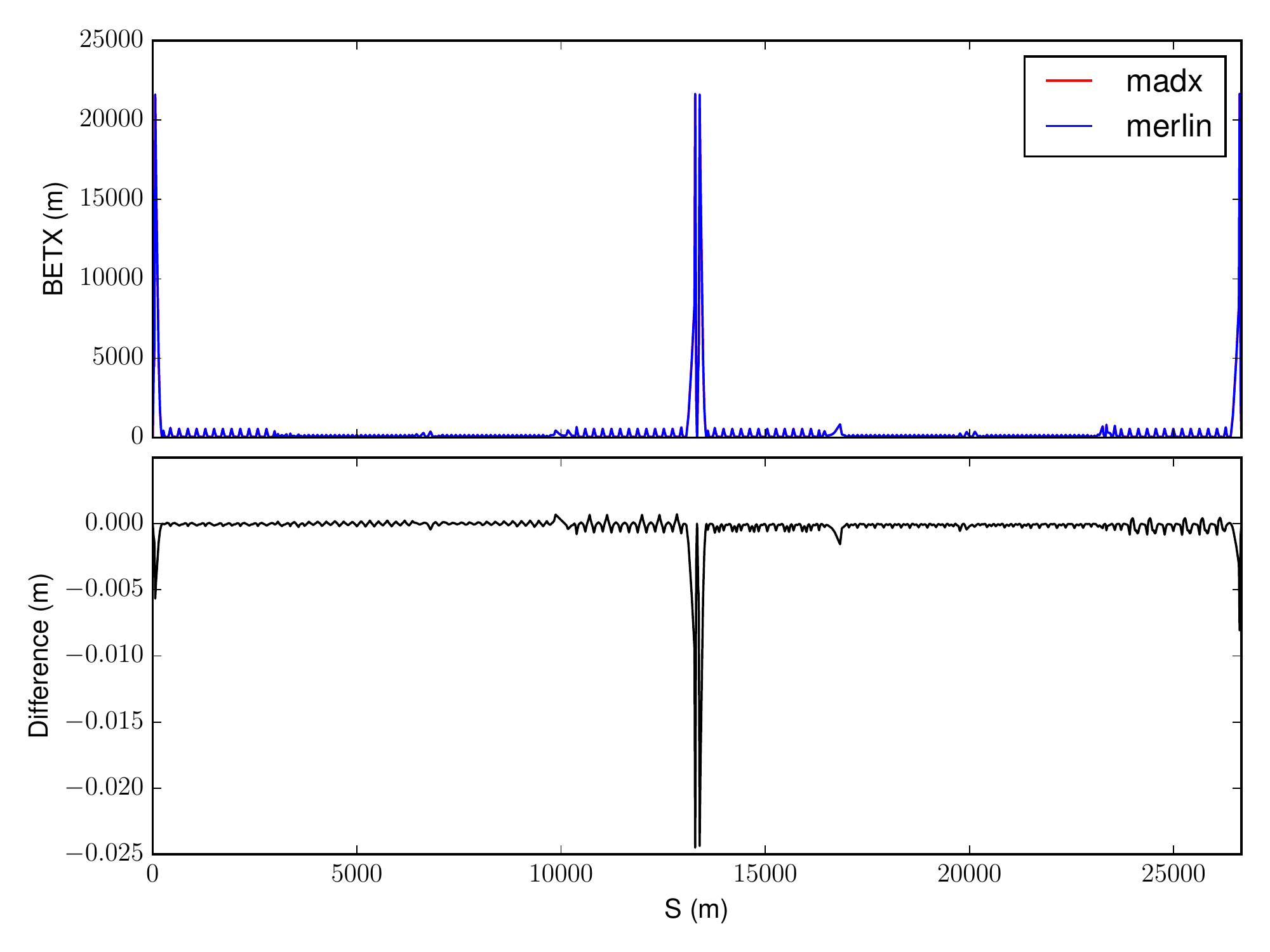}
        \caption{$\beta_x$}
    \end{subfigure}
        \begin{subfigure}[b]{0.49\columnwidth}
        \includegraphics[width=\columnwidth]{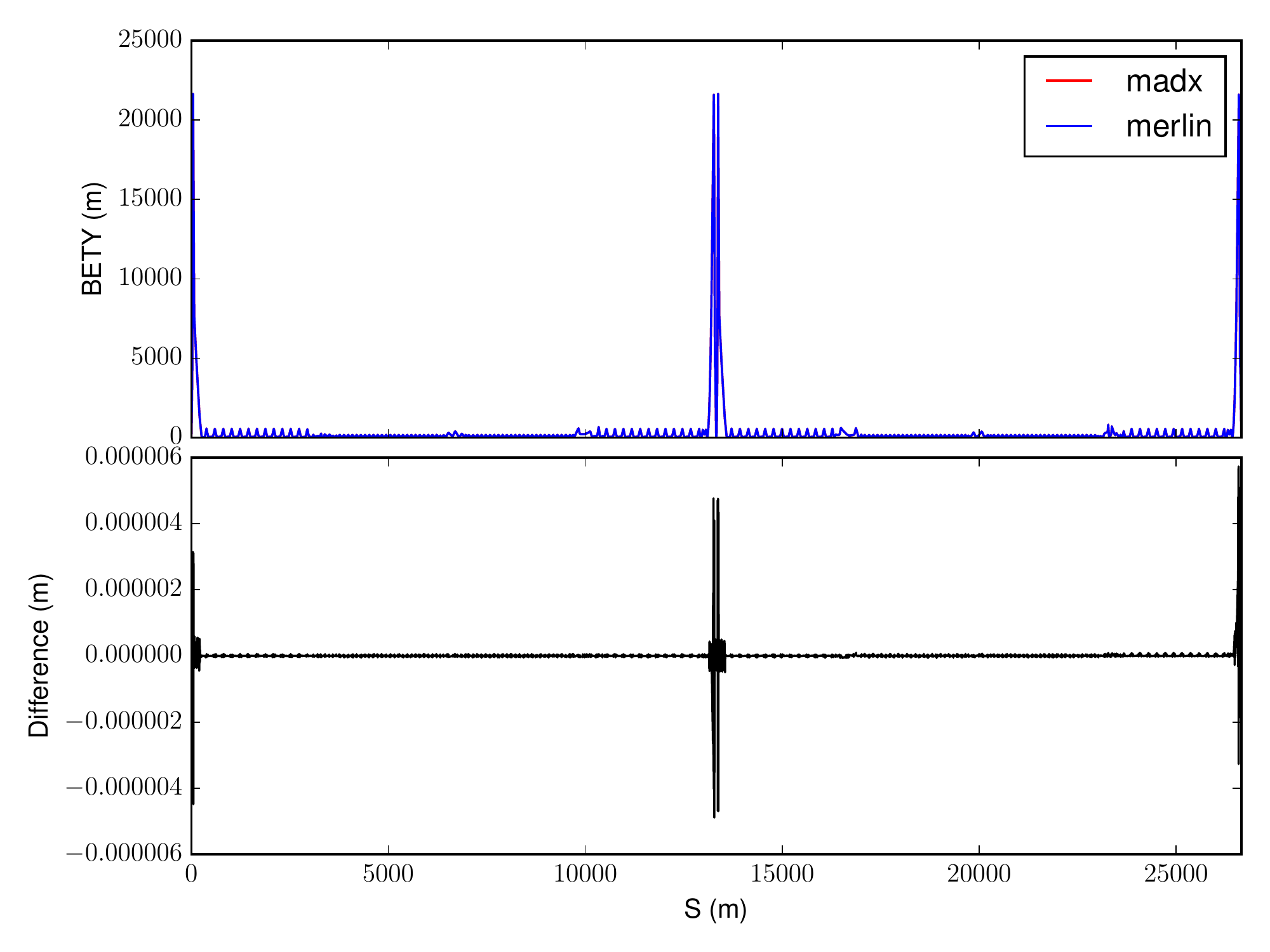}
        \caption{$\beta_y$}
    \end{subfigure}
    
        \begin{subfigure}[b]{0.49\columnwidth}
        \includegraphics[width=\columnwidth]{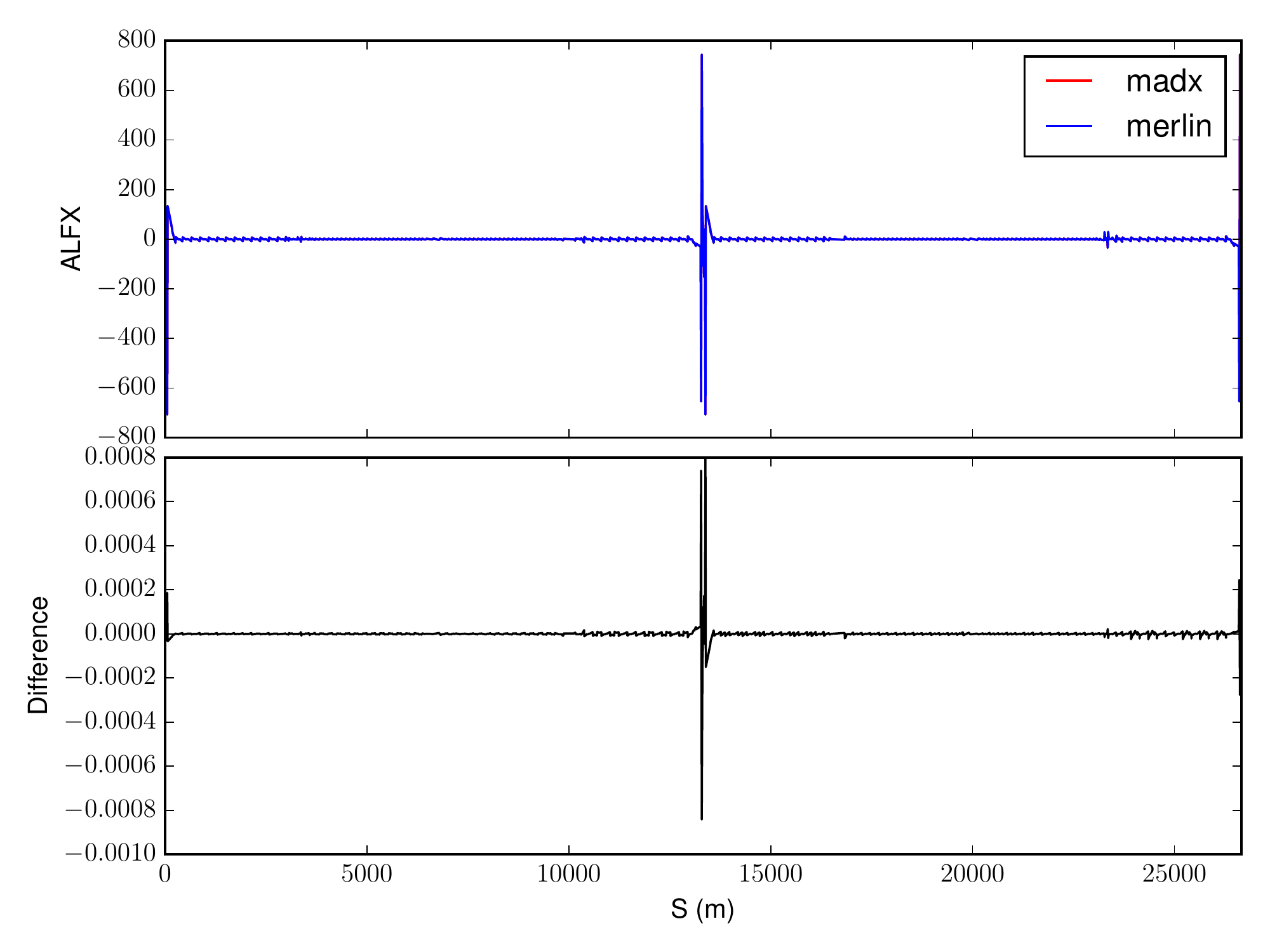}
        \caption{$\alpha_x$}
    \end{subfigure}
        \begin{subfigure}[b]{0.49\columnwidth}
        \includegraphics[width=\columnwidth]{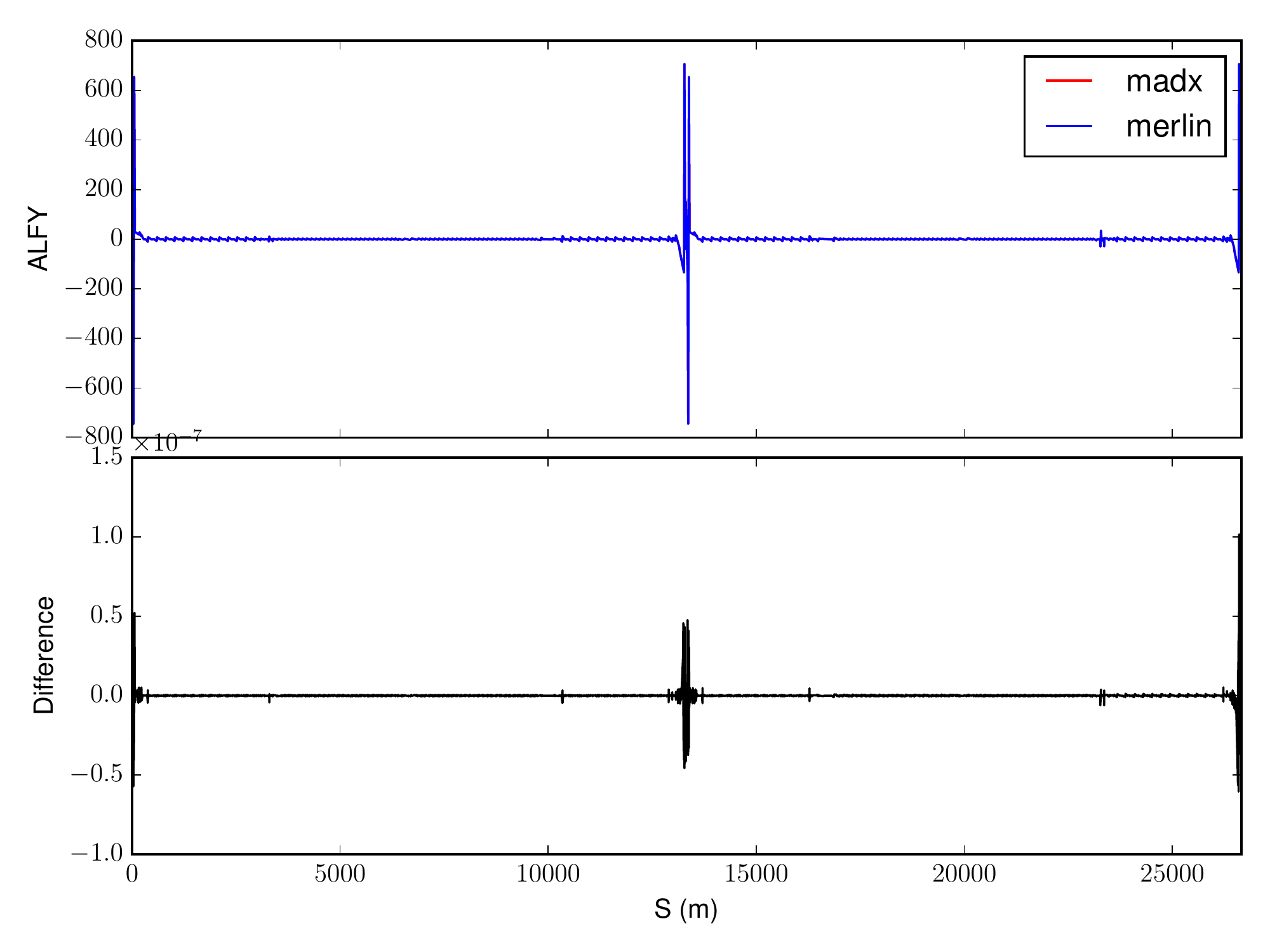}
        \caption{$\alpha_y$}
    \end{subfigure}
    
    \caption{Difference between MERLIN and MAD-X for $\beta$ and $\alpha$-functions for full ring 15 cm round optics.}
    \label{fig:beta_disp_ring_BETX_BETY}
\end{figure}

The optics functions found by MERLIN using tracking can be compared to those found using MAD-X's matrix methods. This is useful to validate the tracking model in MERLIN. Figure \ref{fig:beta_disp_ring_BETX_BETY} shows the $\beta$ and $\alpha$ functions from the two codes, and the difference between them. The greatest difference is in the inner triplets where the $\beta$ function deviates by 25~mm in 20~km.

\subsection{Collimation}
\label{sec:col}

The \texttt{Collimator\ab{}Database} is used to construct collimator apertures using an input file. The jaw half gaps, rotation angle, tilt, and material are defined in the input file and set accordingly when read by the \texttt{Collimator\ab{}Database} class.

The modular approach to collimation allows the user to override the definitions of any aspect. 

For optimisation the \texttt{Cross\ab{}Sections} class calculates and stores all cross sections for a given material. These cross sections are called by the \texttt{Scattering\ab{}Process} classes when performing point like scattering, and in \texttt{ScatteringModel::PathLength()} to retrieve the total mean free path $\lambda_{tot}$. By using this class to compute and save the cross sections, MERLIN minimises computation time at the cost of an inexpensive amount of memory. \texttt{Cross\ab{}Sections} stores the advanced \texttt{ppElasticScatter} and \texttt{ppDiffractiveScatter} classes (for more details on these classes see~\cite{molson}), allowing access to them during the collimation processes.

\texttt{ScatteringProcess} is an abstract base class for individual point-like scattering processes. It contains a pointer to the \texttt{Material} and \texttt{CrossSections} classes, the process cross section, the beam energy, and two functions: \texttt{Configure()} and \texttt{Scatter()}.

\begin{figure}[!htbp]
	\begin{center}
		\includegraphics[width=0.6\linewidth]{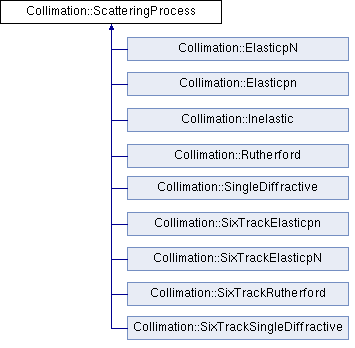}
		\caption{Scattering processes currently available in MERLIN}
		\label{fig:scatpro}
	\end{center}
\end{figure}

MERLIN contains a number of \texttt{ScatteringProcess}es, including the SixTrack-like variants (based on K2 scattering); those currently available are shown in Fig.~\ref{fig:scatpro}.

The user can select from several predefined \texttt{ScatteringModel}s. \texttt{ScatteringModelMerlin} provides the full scattering physics as described in \cite{practicalpomeron}. \texttt{ScatteringModelSixTrack} provides a model based on the K2 scattering found in SixTrack. \texttt{ScatteringModelSixTrackIoniz}, \texttt{ScatteringModelSixTrackElastic}, \texttt{ScatteringModelSixTrackSD} provide hybrid models useful for testing the individual scattering processes independently. Table~\ref{tab:ScatProCombi} shows combinations of processes in each model. It is also possible for the user to create a \texttt{ScatteringModel} and customise it by adding their required processes.

\begin{table}[!htbp]
   \centering
   \caption{Preset combinations of \texttt{ScatteringProcess}es and ionisation in MERLIN. ST refers to the SixTrack-like process, M to the MERLIN process, and all combinations include an inelastic process.}
   \begin{tabular}{lccccc}
       \hline
       \textbf{Process} & \textbf{SixTrack}     & \textbf{SixTrackElastic}	&\textbf{SixTrackSD}	&\textbf{SixTrackIoniz} & \textbf{Merlin} \\
       \hline
Rutherford	      	 &ST &ST &ST &ST & M   	\\
pn Elastic           &ST &M &ST &ST & M   	\\
pN Elastic           &ST &M &ST &ST & M  	\\
Single Diffractive   &ST &ST &M &ST & M   	\\
Ionisation           &ST &ST &ST &M & M  	\\
       \hline
   \end{tabular}
   \label{tab:ScatProCombi}
\end{table}

The \texttt{Scattering\ab{}Model} class contains the functions required for performing collimation. A predefined or user-created combination of \texttt{Scattering\ab{}Processes} may be used, and are handled by the \texttt{Scattering\ab{}Model} in order to compute cross sections, path lengths, and perform bulk (ionisation and MCS) and point-like scattering. The \texttt{Scattering\ab{}Process} must be attached to the \texttt{Collimate\ab{}Proton\ab{}Process}.

The path length is the average distance a proton travels through a material before colliding with a material nucleus, it is calculated in the \texttt{Path\ab{}Length()} function for each proton at each iteration of scattering, using the mean free path.

As a proton travels through a material it collides with electrons, these collisions may result in the removal of electrons, and a loss in proton energy. The interaction is defined by the Bethe-Bloch equation~\cite{PDG}. MERLIN offers an overloaded \texttt{Energy\ab{}Loss()} function to calculate the energy loss that takes place due to ionisation in a material. The basic function performs energy loss according to the Bethe-Bloch equation~\cite{PDG}. The overloaded function is a more complete treatment of the energy loss due to ionisation using higher-order corrections (considered in~\cite{molson}). In summary, MERLIN only adds the effects that are relevant to LHC energies: the effect due to the dielectric polarisability of solid materials, the Mott correction which is an enhancement from close collisions due to spin, and the finite size correction taking into account the size and structure of the proton. As well as these corrections, the energy spread of the outgoing proton is sampled using the Landau distribution, which is a more accurate representation of the physical effect~\cite{PDG}.

A proton travelling through a material will perform many small-angle elastic scatters from the electrons and nuclei, known as multiple Coulomb scattering (MCS), this is performed in the \texttt{Straggle} function.

\texttt{ParticleScatter()} is called when a particle has travelled its path length and remains in the material, which means it will interact with a material nucleus or nucleon (\textit{i.e.} a point-like scatter).

\subsection{Outputs}

\begin{figure}[!htbp]
    \centering
    \begin{subfigure}[b]{0.49\columnwidth}
        \includegraphics[width=\columnwidth]{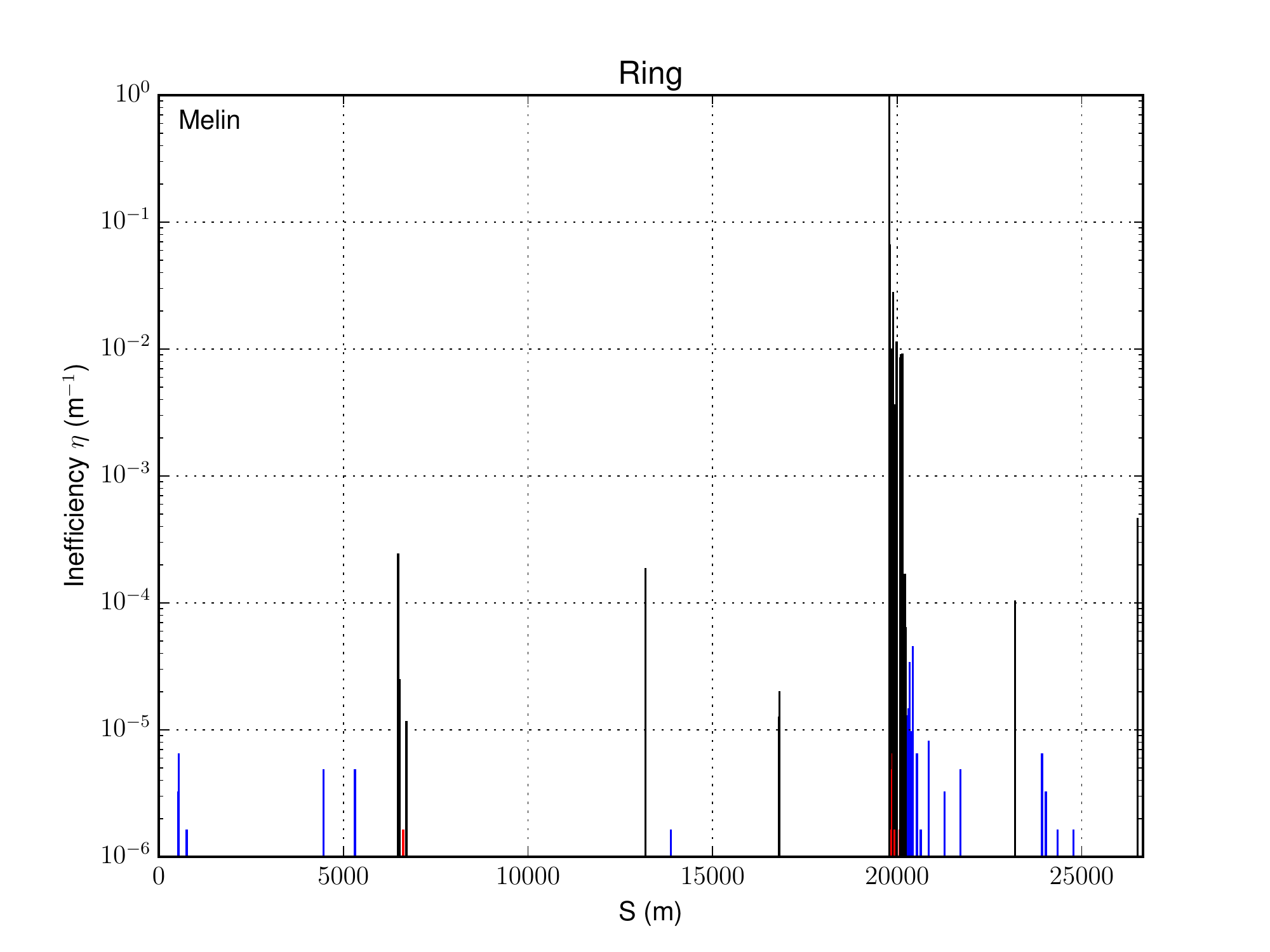}
        \caption{Full LHC ring}
        \label{fig:Ring}
    \end{subfigure}
        \begin{subfigure}[b]{0.49\columnwidth}
        \includegraphics[width=\columnwidth]{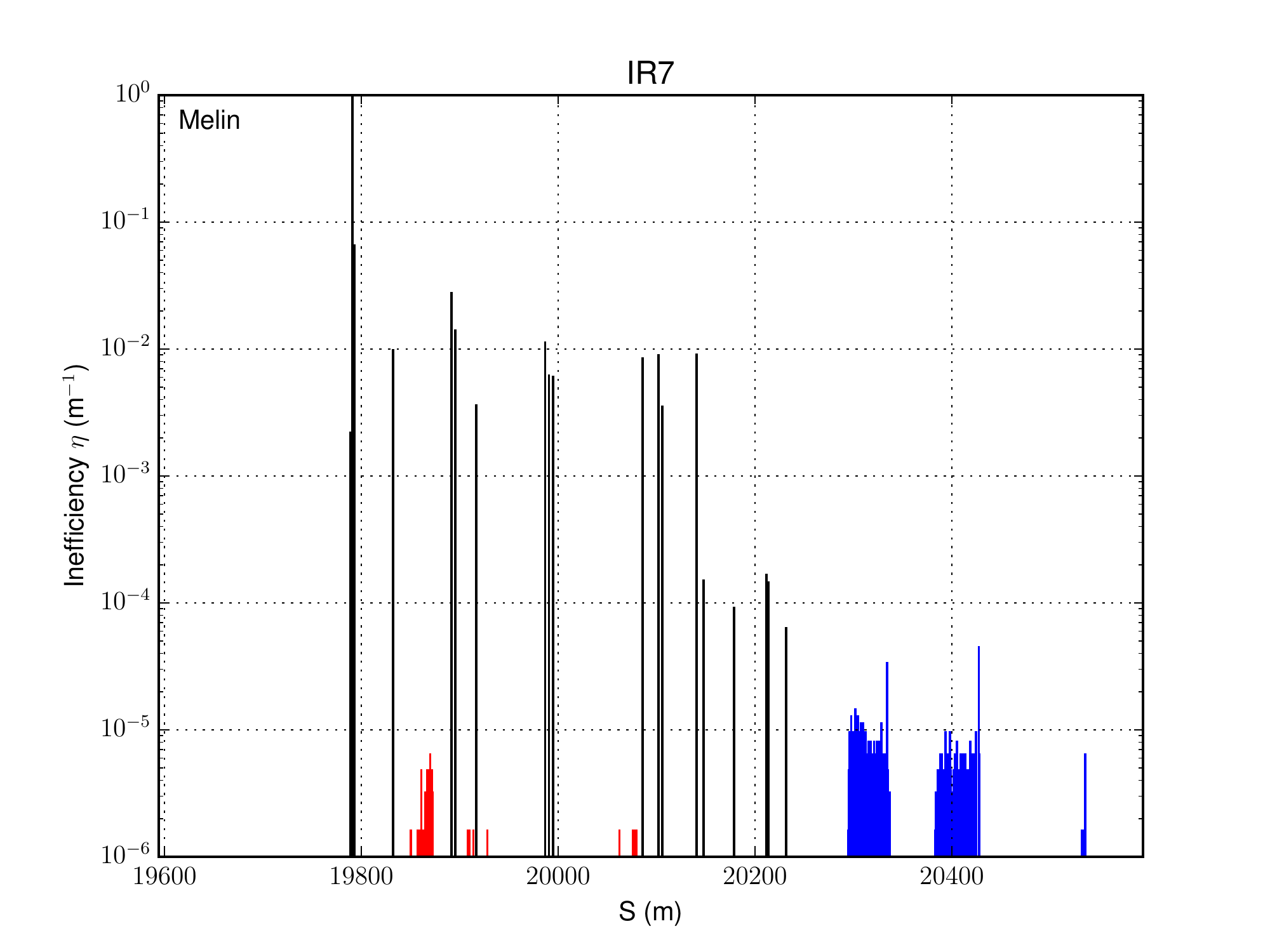}
        \caption{Collimation region IR7}
        \label{fig:IR7}
    \end{subfigure}
    \caption{Loss maps on a log scale for LHC at 6.5~TeV squeezed settings. Black, red and blue show collimators, warm and cold elements respectively.}
    \label{fig:lossmap}
\end{figure}

The main output from a collimation simulation is the loss map. This is a count of particle losses binned either by element or by position around the ring. MERLIN records these using the \texttt{CollimationOutput} class. Typically, losses from collimators and other elements are recorded together. No significant post processing is needed, as all the aperture checking happens on-line. Figure \ref{fig:lossmap} shows a loss map for the LHC ring at the 6.5~TeV squeezed configuration. The losses in the betatron collimation region in IR7 are clearly visible, as are losses in the TCTs around the experiments.

\begin{figure}[!htbp]
	\begin{center}
		\includegraphics[width=1\linewidth]{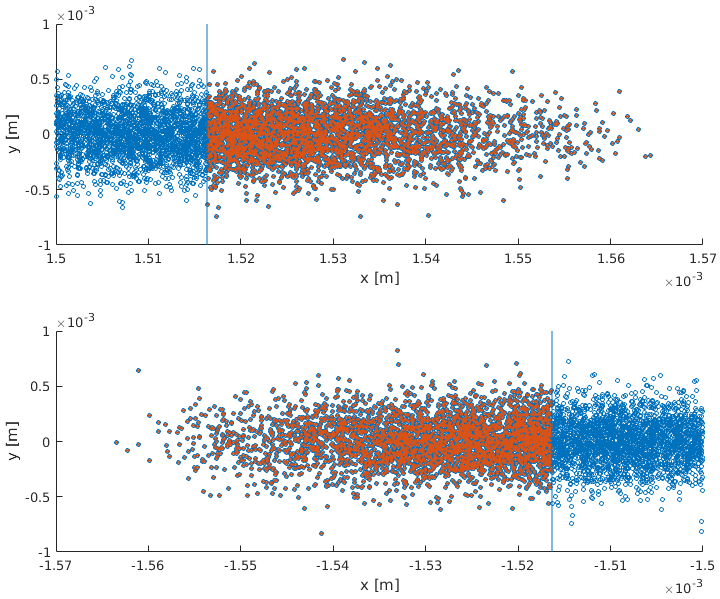}
		\caption{Initial distribution (blue) and impacts recorded on the primary horizontal collimator (orange) using \texttt{JawImpact}, for the positive (above) and negative (below) collimator jaws. The blue line indicates the collimator aperture, where the jaw begins. This simulation is for the 6.5 TeV LHC at flat top, using beam 2.}
		\label{fig:jawimpact}
	\end{center}
\end{figure}

MERLIN also provides outputs that are useful for diagnostics and better understanding of the collimation process. The \texttt{Scattering\ab{}Model::Jaw\ab{}Impact()} function outputs the coordinates of particles that impact the front face of selected collimators. As well as coordinates, the turn at which the impact occurred is also output, allowing the user to observe any change in the impact parameter over time. This output is useful for observing the effect of the HEL. An example of this output for a single turn in a collimation simulation is shown in Fig.~\ref{fig:jawimpact}, showing the correspondence between the initial distribution (which starts immediately in front of the primary collimator), with the recorded impact coordinates.

\begin{figure}[!htbp]
	\begin{center}
		\includegraphics[width=1\linewidth]{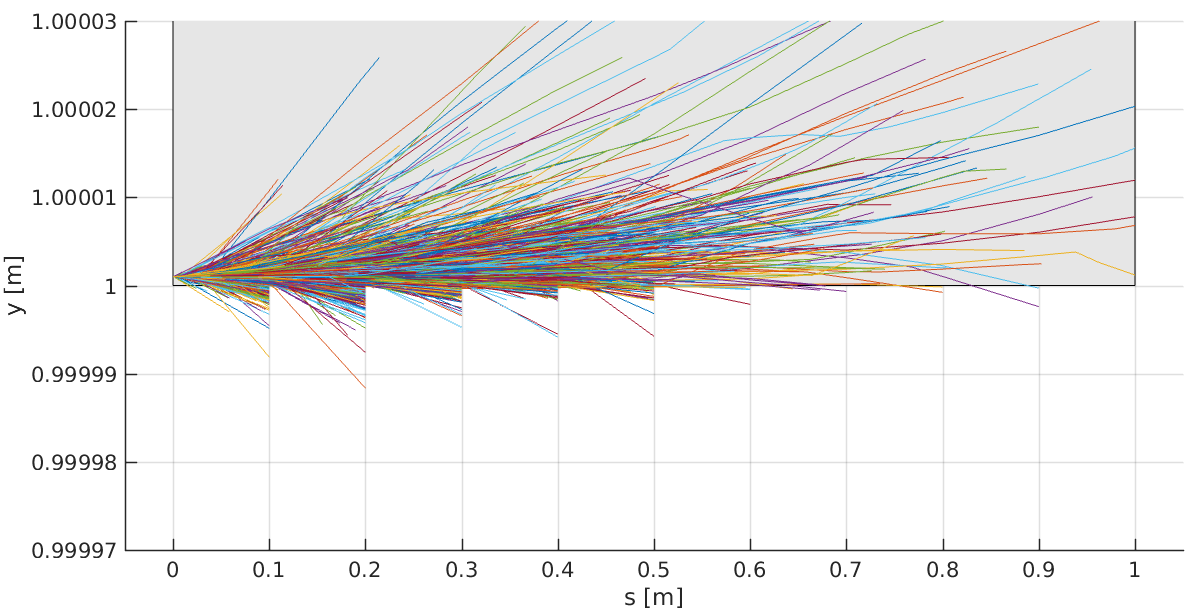}
		\caption{\texttt{ScatterPlot()} output showing the proton tracks taken in a 1~m long copper collimator with an impact parameter of 1~$\mu$m in the $y$ plane, using a 10~cm collimation bin size. The grey area indicates the collimator jaw, and the particles are not tracked by this output if they exit the collimator jaw.}
		\label{fig:scatterplot}
	\end{center}
\end{figure}

The \texttt{Scattering\ab{}Model::Scatter\ab{}Plot()} function stores the position of particles at each path length step in order to plot scattering tracks along the collimator. This function has been a useful tool for debugging the collimation process, ensuring that aperture checks are performed at appropriate intervals, and showing the effect of the collimation bin size. In Fig.~\ref{fig:scatterplot} the effect of the collimation bin size is illustrated. Particles undergo scattering, MCS, and ionisation energy loss for as long as they are in the collimator jaw. Even if a particle has exited a collimator jaw, the aperture check cannot take place until the particle has travelled a path length $l_{path}$, or at the end of the collimation bin. This is evident as particle tracks abruptly stop at 10~cm intervals in the figure. By reducing the bin size, a small path length is forced, and computation time will increase, however by using a larger bin size protons may undergo significantly more bulk scattering (MCS and ionisation energy loss) when they have in fact already left the collimator jaw. The 10~cm bin size that is used by default allows regular aperture checks without enforcing too small a path length, or compromising the condition of protons that return to the bunch after undergoing scattering in a collimator jaw.

\begin{figure}[!htbp]
	\begin{center}
		\includegraphics[width=1\linewidth]{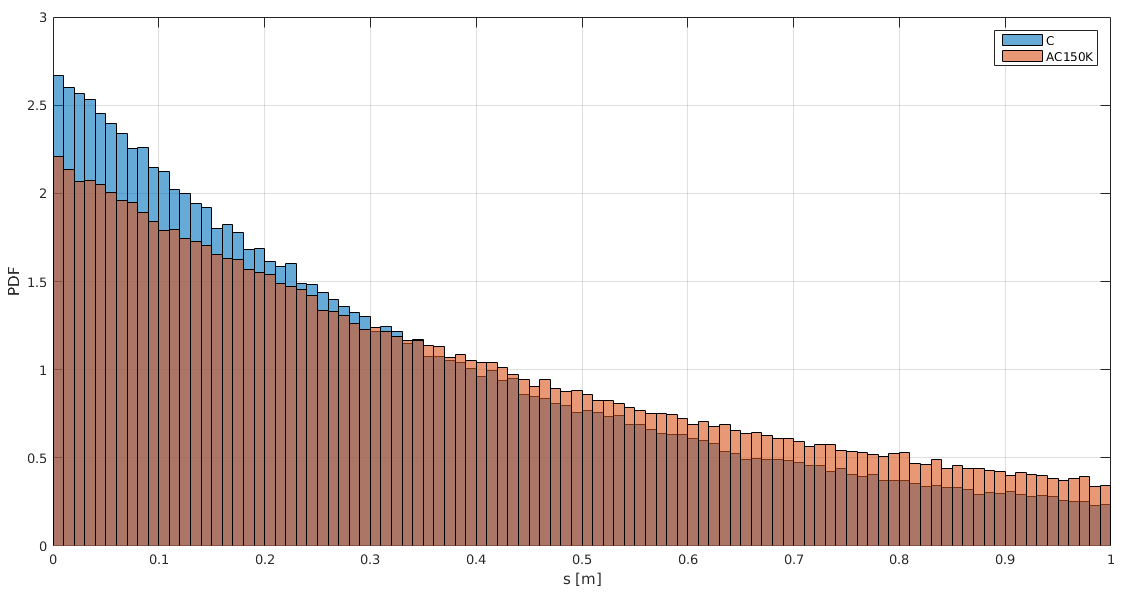}
		\caption{Inelastic proton interactions (proton losses) in a secondary collimator in the nominal LHC using the \texttt{JawInelastic} output. Comparing losses in pure carbon (blue) with CFC AC150K (orange).}
		\label{fig:JawInelastic}
	\end{center}
\end{figure}

\texttt{Scattering\ab{}Model::Jaw\ab{}Inelastic()} stores the coordinates of inelastic interactions. This provides a necessary comparison tool to observe the effect of different collimator materials; an example histogram of the distribution of losses in a given collimator is shown in Fig.~\ref{fig:JawInelastic}.

\begin{figure}[!htbp]
	\begin{center}
		\includegraphics[width=1\linewidth]{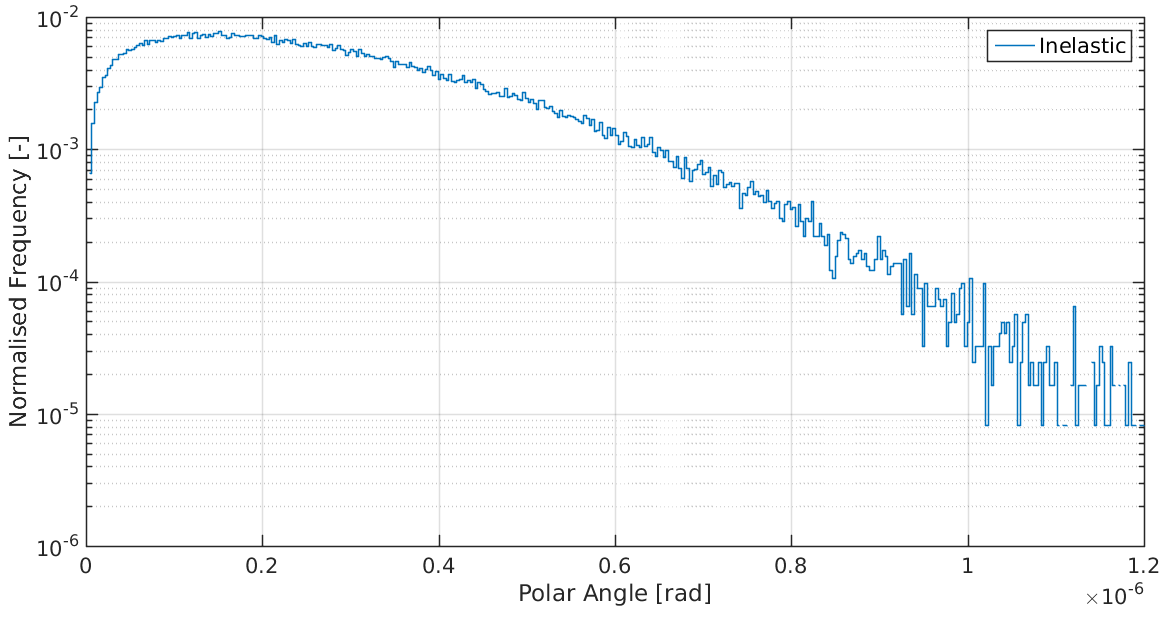}
		\caption{Polar angle histogrammed in the \texttt{Output\ab{}Select\ab{}Scatter\ab{}Histogram()} output function, showing the angular distribution of particles that have undergone inelastic interactions in a collimator made of CFC AC150K$^{\textregistered}$. In reality this angular spread is given by MCS.}
		\label{fig:selectscatter}
	\end{center}
\end{figure}

\texttt{Scattering\ab{}Model::Selec\ab{}Scatter()} stores the coordinates of selected interactions, outputting the momentum transfer $t$ and the polar angle $\theta$ as well as other quantities. As the raw data files produced can be very large, the \texttt{Output\ab{}Select\ab{}Scatter\ab{}Histogram()} function was created to histogram the data and produce a smaller output. An example of the histogrammed polar angle data is shown in Fig.~\ref{fig:selectscatter}.

\section{Development}

MERLIN consists of 39k lines of C++ code (LOC), along with 4.8k LOC of examples and 1.8 LOC of tests. CMake is used as the build system, allowing it to be compiled on a range of platforms. The source code is held in Git revision control repository~\cite{merlingit}.

MERLIN has a growing automated test suite. This is run daily on several physical and virtual machines with a range of operating systems, compilers and CPU architectures. The results uploaded to the CERN CDash server~\cite{cdash}. This allows rapid identification of issues in new development. The test suite performs dynamic analysis using the Vagrind toolset~\cite{valgrind} to identify memory leaks and other related errors.

The Git distributed revision system offers a systematic method for making and recording changes to the source code. New features and fixes can be developed on a branch, and then merged into the master branch when tested and ready. This keeps the master branch in a usable condition and makes it easy to back out a change if it is found to have a negative effect. The master branch is hosted on the GitHub service, which also provides issue tracking and management of branches and merges.

\section{Acknowledgements}

This work is supported by STFC (UK) grant High Luminosity LHC : UK (HL-LHC-UK), grant number ST/N001621/1.

\end{document}